# ANALYTIC SOLUTIONS OF THE SCHRÖDINGER EQUATION WITH NON-CENTRAL GENERALIZED INVERSE QUADRATIC YUKAWA POTENTIAL

C. O. Edet[1], P. O Okoi[2] and S. O Chima[2]

[1]Theoretical Physics Group, Department of Physics, University of Port Harcourt, Choba, Nigeria

[2]Department of Physics, University of Calabar, Nigeria

E-mail for correspondence: collinsokonedet@gmail.com

## ABSTRACT

This study presents the solutions of Schrödinger equation for the Non-Central Generalized Inverse Quadratic Yukawa Potential within the framework of Nikiforov-Uvarov. The radial and angular part of the Schrödinger equation are obtained using the method of variable separation. More so, the bound states energy eigenvalues and corresponding eigenfunctions are obtained analytically. Numerical results were obtained for the Generalized Inverse Quadratic Yukawa Potential for comparison sake. It was found out that our results agree with existing literature.

## 1 INTRODUCTION

Physicists over the years have developed strong interest in searching for the solution of the Schrödinger equation with some potentials [1-7]. This is because, finding the analytical solution of the Schrödinger equation is extremely crucial in nonrelativistic quantum mechanics and the eigenfunction contains all the necessary information required to describe a quantum system under consideration.

The analytical solution of the Schrodinger equation with $\ell = 0$ and $\ell \neq 0$ for some potentials (central and non-central) has been addressed by many researchers in nonrelativistic quantum mechanics and relativistic quantum mechanics for bound and scattering states problems[8-16]. Some of the potentials addressed in these studies include; a class of Yukawa potential [17], Coulomb ring-shaped potential [18], ring-shaped Woods-Saxon potential [19], Hartman potential [20-21], Coulombic ring-shaped potential[22] double ring-shaped oscillator potential[23] ring-shaped non-spherical harmonic oscillator potential[24-25], spherically harmonic oscillatory ring-shaped potential [26-27], Poschl Teller double-ring-shaped Coulomb potential [28], Manning-Rosen potentials[29-31], inversely quadratic Yukawa potential[32] and Eckart[33]. In most of these studies, the authors used certain well known approximations to overcome the centrifugal barrier ($\ell \neq 0$) and obtained the eigenvalues and eigenfunctions using different methods.

The methods which have been employed to solve the differential equation arising from these considerations include; the asymptotic iteration method (AIM) [34-36], exact quantization rules [37-38], Nikiforov–Uvarov (NU) method [39-41], modified factorization method [42-43], supersymmetric quantum mechanics (SUSYQM) [44-55], and the functional analysis approach (FAA) [46-47].

The generalized inverse quadratic Yukawa potential (GIQYP) was first proposed by [48-49]. This potential is a superposition of the inverse quadratic Yukawa (IQY) [32] and the Yukawa potential [50]. It is asymptotic to a finite value as $r \to \infty$ and becomes infinite at $r = 0$. [51] solved this potential within the framework of the proper quantization rule and eigenfunction was obtained via the Formula Method [52]. The generalized inverse quadratic Yukawa potential model is of the form [51]

$$V(r) = -V_1 \left(1 + \frac{e^{-\alpha r}}{r}\right)^2 \quad (1)$$

[48] Compared the behaviour of the Yukawa-type potential with the Yukawa potential and the IQY potential for screening parameter values, it was noted that differences do not exist between these three potentials.

The Ring shaped potential have a wide range of applications in quantum chemistry and nuclear physics [53]. They have very important role in describing ring-shaped molecules like benzene and the interactions between deformed pair of nuclei [54-55]. They have also been used in demonstrating some of the pseudospin symmetry in nuclei physics [27]. The exact results can be used in accounting for some axial symmetric system in quantum chemistry. [26] proposed a non-central potential as;

$$V(\theta) = \frac{\hbar^2}{2\mu r^2}\left(\frac{C + B\cos^2\theta + A\cos^4\theta}{\sin^2\theta \cos^2\theta}\right) \quad (2)$$

Motivated by this numerous studies, we attempt to propose a non-central generalized inverse quadratic Yukawa potential (GIQYP) by selecting $V(r)$ as the generalized inverse quadratic Yukawa potential (GIQYP). The non-central generalized inverse quadratic Yukawa potential (GIQYP) is composed of generalized inverse quadratic Yukawa potential (GIQYP) plus a Novel Angle Dependent (NAD) potential. It can be written as

$$V(r,\theta) == -\frac{\eta e^{-2\alpha r}}{r^2} - \frac{\upsilon e^{-\alpha r}}{r} - \mathcal{H} + \frac{\hbar^2}{2\mu r^2}\left(\frac{C+B\cos^2\theta + A\cos^4\theta}{\sin^2\theta \cos^2\theta}\right) \quad (3)$$

Where $\eta = \mathcal{H} = V_1$ and $\upsilon = 2V_1$

The Generalized inverse quadratic Yukawa potential reduces to a constant potential when $\eta = \upsilon = 0$.

The organization of the work is as follows: In the next section, we give a review of the NU method. In Sect. 3, this method is applied to obtain the bound state solutions with Non-Central Generalized Inverse Quadratic Yukawa Potential. In Sect. 4, we obtain numerical results while in Sect. 5, we give a brief concluding remark.

## 2. REVIEW OF NIKIFOROV-UVAROV METHOD

The Nikiforov-Uvarov (NU) method is based on solving the hypergeometric-type second-order differential equations by means of the special orthogonal functions [56-57]. The main equation which is closely associated with the method is given in the following form [58];

$$\psi''(z) + \frac{\tilde{\tau}(z)}{\sigma(z)}\psi'(z) + \frac{\tilde{\sigma}(z)}{\sigma^2(z)}\psi(z) = 0 \quad (4)$$

Where $\sigma(z)$ and $\tilde{\sigma}(z)$ are polynomials at most second-degree, $\tilde{\tau}(z)$ is a first-degree polynomial and $\psi(z)$ is a function of the hypergeometric-type.

The exact solution of Eq. (4) can be obtained by using the transformation

$$\psi(z) = \phi(z)y(z) \quad (5)$$

This transformation reduces Eq. (4) into a hypergeometric-type equation of the form

$$\sigma(z)y''(z) + \tau(z)y'(z) + \lambda y(z) = 0 \quad (6)$$

The function $\phi(z)$ can be defined as the logarithm derivative

$$\frac{\phi'(z)}{\phi(z)} = \frac{\pi(z)}{\sigma(z)} \quad (7)$$

where $\pi(z) = \frac{1}{2}[\tau(z) - \tilde{\tau}(z)]$ \quad (8)

with $\pi(z)$ being at most a first-degree polynomial. The second $\psi(z)$ being $y_n(z)$ in Eq. (5), is the hypergeometric function with its polynomial solution given by Rodrigues relation

$$y^{(n)}(z) = \frac{B_n}{\rho(z)} \frac{d^n}{ds^n}[\sigma^n(z)\rho(z)] \qquad (9)$$

Here, $B_n$ is the normalization constant and $\rho(z)$ is the weight function which must satisfy the condition

$$(\sigma(z)\rho(z))' = \sigma(z)\tau(z) \qquad (10)$$

$$\tau(z) = \tilde{\tau}(z) + 2\pi(z) \qquad (11)$$

It should be noted that the derivative of $\tau(s)$ with respect to $s$ should be negative. The eigenfunctions and eigenvalues can be obtained using the definition of the following function $\pi(s)$ and parameter $\lambda$, respectively:

$$\pi(z) = \frac{\sigma'(z) - \tilde{\tau}(z)}{2} \pm \sqrt{\left(\frac{\sigma'(z) - \tilde{\tau}(z)}{2}\right)^2 - \tilde{\sigma}(z) + k\sigma(z)} \qquad (12)$$

where $k = \lambda - \pi'(z)$ \qquad (13)

The value of $k$ can be obtained by setting the discriminant of the square root in Eq. (12) equal to zero. As such, the new eigenvalue equation can be given as

$$\lambda_n = -n\tau'(z) - \frac{n(n-1)}{2}\sigma''(z), \quad n = 0,1,2,\ldots \qquad (14)$$

## 3 SEPARATION OF VARIABLES FOR THE SCHRODINGER EQUATION

In spherical coordinates $(r, \theta, \phi)$, the Schrodinger equation with potentials $V(r,\theta)$, respectively, can be written as follows[26]:

$$-\frac{\hbar^2}{2\mu}\nabla^2 \psi(r,\theta,\phi) + V(r,\theta)\psi(r,\theta,\phi) = E\psi(r,\theta,\phi) \qquad (15)$$

where $E$ is the non-relativistic energy of the system, $\mu$ denotes the rest mass of the particle and $\hbar$ is the planck constant. The Schrodinger equation with potential is given by[54];

$$\left[-\frac{\hbar^2}{2\mu}\left[\frac{1}{r^2}\frac{\partial}{\partial r}r^2\frac{\partial}{\partial r} + \frac{1}{r^2 \sin\theta}\frac{\partial}{\partial \theta}\left(\sin\theta \frac{\partial}{\partial \theta}\right) + \frac{1}{r^2 \sin^2\theta}\frac{\partial^2}{\partial \phi^2}\right] + V(r,\theta) - E\right]\psi(r,\theta,\phi) = 0 \qquad (16)$$

$$\psi(r,\theta,\phi) = R(r)\Theta(\theta)\Phi(\phi) \qquad (17)$$

Substituting Eq. (3) into Eq.(16), we have

$$\left[-\frac{\hbar^2}{2\mu}\left[\frac{1}{r^2}\frac{\partial}{\partial r}r^2\frac{\partial}{\partial r}+\frac{1}{r^2\sin\theta}\frac{\partial}{\partial\theta}\left(\sin\theta\frac{\partial}{\partial\theta}\right)+\frac{1}{r^2\sin^2\theta}\frac{\partial^2}{\partial\phi^2}\right]+\left(-\frac{Ae^{-2\alpha r}}{r^2}-\frac{Be^{-\alpha r}}{r}-C+\frac{\hbar^2}{2\mu r^2}\left(\frac{C+B\cos^2\theta+A\cos^4\theta}{\sin^2\theta\cos^2\theta}\right)\right)-E\right]\psi(r,\theta,\phi)=0$$
(18)

Substituting Eq. (17) into Eq. (18) and using the standard procedure of separating variables, we obtain the following differential equations:

$$\frac{d^2R_{nl}}{dr^2}+\left[\frac{2\mu E_{nl}}{\hbar^2}-\frac{2\mu}{\hbar^2}\left(-\frac{Ae^{-2\alpha r}}{r^2}-\frac{Be^{-\alpha r}}{r}-C\right)-\frac{\Lambda}{r^2}\right]R_{nl}(r)=0 \qquad (19)$$

$$\frac{d^2\Theta(\theta)}{d\theta^2}+\frac{\cos\theta}{\sin\theta}\frac{d\Theta(\theta)}{d\theta}+\left(\Lambda-\left(\frac{\hbar^2}{2\mu r^2}\left(\frac{C+B\cos^2\theta+A\cos^4\theta}{\sin^2\theta\cos^2\theta}\right)\right)-\frac{m^2}{\sin^2\theta}\right)\Theta(\theta)=0 \qquad (20)$$

$$\frac{d^2\Phi(\phi)}{d\phi^2}+m^2\Phi(\phi)=0 \qquad (21)$$

where $m^2$ and $\Lambda$ are separation constants, which are real and dimensionless. The solution of Eq. (21) is periodic and for bound state $\Phi(\phi)$ satisfies the periodic boundary condition $\Phi(\phi+2\pi)$ and its solutions become,

$$\Phi(\phi)=\frac{1}{\sqrt{2\pi}}e^{-im\phi}, m=0,\pm1,\pm2,\ldots \qquad (22)$$

## 3.1 Solutions of the radial Schrodinger equation for Non-Central Generalized Inverse Quadratic Yukawa Potential

$$\frac{d^2R_{nl}}{dr^2}+\left[\frac{2\mu E_{nl}}{\hbar^2}-\frac{2\mu}{\hbar^2}\left(-\frac{V_1 e^{-2\alpha r}}{r^2}-\frac{2V_1 e^{-\alpha r}}{r}-V_0\right)-\frac{\Lambda}{r^2}\right]R_{nl}(r)=0 \qquad (23)$$

The radial part of the Schrödinger equation for this potential can be solved exactly for $l=0$ (s-wave) but cannot be solved for this potential for $l\neq 0$. To obtain the solution for $l\neq 0$, we employ the approximation scheme proposed by Greene and Aldrich [59] to deal with the centrifugal term, which is given as;

$$\frac{1}{r^2}\approx\frac{\alpha^2}{(1-e^{-\alpha r})^2} \qquad (24)$$

It is noted that for a short-range potential, the relation (eqs. 24 ) is a good approximation to $\frac{1}{r^2}$, as proposed by Greene and Aldrich [59,60-61]. The implies that Eq. (24) is not a good approximation to the centrifugal barrier when the screening parameter $\alpha$ becomes large. Thus, the approximation is valid when $\alpha \ll 1$. Substituting the approximation (Eq.24) into Eq. (23), we obtain an equation of the form;

$$\frac{d^2R_{nl}}{dr^2} + \left[\frac{2\mu E_{nl}}{\hbar^2} - \frac{2\mu}{\hbar^2}\left(-\frac{A\alpha^2 e^{-2\alpha r}}{(1-e^{-\alpha r})^2} - \frac{B\alpha e^{-\alpha r}}{(1-e^{-\alpha r})} - C\right) - \frac{\Lambda\alpha^2}{(1-e^{-\alpha r})^2}\right]R_{nl}(r) = 0 \quad (25)$$

Eq. (25) can be simplified into the form and introducing the following dimensionless abbreviations

$$\begin{cases} \varepsilon_n = \frac{2\mu(E_{nl}+V_1)}{\hbar^2\alpha^2} \\ \beta = \frac{2\mu V_1}{\hbar^2} \\ \chi = \frac{4\mu V_1}{\hbar^2\alpha} \end{cases} \quad (26)$$

Using a transformation $z = e^{-\alpha r}$ so as to enable us apply the NU method as a solution of the hypergeometric type

$$\frac{d^2R_{n\ell}(r)}{dr^2} = \alpha^2 z^2 \frac{d^2R_{n\ell}(z)}{dz^2} + \alpha^2 z \frac{dR_{n\ell}(z)}{dz} \quad (28)$$

$$\frac{d^2R_{n\ell}}{dz^2} + \frac{(1-z)}{z(1-z)}\frac{dR_{n\ell}}{dz} + \frac{1}{z^2(1-z)^2}[-\varepsilon_n(1-z)^2 + \beta z^2 + \chi z(1-z) - \Lambda]R_{n\ell}(z) = 0 \quad (27)$$

We obtain the differential equation

$$\frac{d^2R_{n\ell}}{dz^2} + \frac{(1-z)}{z(1-z)}\frac{dR_{n\ell}}{dz} + \frac{1}{z^2(1-z)^2}[-(\varepsilon_n - \beta + \chi)z^2 + (2\varepsilon_n + \chi)z - (\varepsilon_n + \Lambda)]R_{n\ell}(z) = 0 \quad (29)$$

Comparing Eq. (29) and Eq. (4), we have the following parameters

$$\begin{cases} \tilde{\tau}(z) = 1 - z \\ \sigma(z) = z(1-z) \\ \tilde{\sigma}(s) = -(\varepsilon_n - \beta + \chi)z^2 + (2\varepsilon_n + \chi)z - (\varepsilon_n + \Lambda) \end{cases} \quad (30)$$

Substituting these polynomials into Eq. (12), we get $\pi(s)$ to be

$$\pi(z) = -\frac{z}{2} \pm \sqrt{\left(\frac{1}{4} + (\varepsilon_n - \beta + \chi) - k\right)z^2 + (-(2\varepsilon_n + \chi) + k)z + (\varepsilon_n + \Lambda)} \quad (31)$$

further rearranged as;

$$\pi(z) = -\frac{z}{2} \pm \sqrt{\left(\frac{1}{4} + \varepsilon_n - \beta + \chi - k\right)z^2 + (k - 2\varepsilon_n - \chi)z + (\varepsilon_n + \Lambda)} \quad (32)$$

To find the constant $k$, the discriminant of the expression under the square root of Eq. (31) should be equal to zero. As such, we have that

$$k_\pm = -(2\Lambda - \chi) \pm 2\sqrt{\varepsilon_n + \Lambda}\sqrt{\frac{1}{4} + \Lambda - \beta} \quad (33)$$

Substituting Eq. (33) into Eq. (31) yields

$$\pi = -\frac{z}{2} \pm \begin{cases} (\sqrt{\eta_1} - \sqrt{\eta_3})z - \sqrt{\eta_1}; & for\ k_+ = -(\eta_2) + 2\sqrt{\eta_1}\sqrt{\eta_3} \\ (\sqrt{\eta_1} - \sqrt{\eta_3})z + \sqrt{\eta_1}; & for\ k_- = -(\eta_2) - 2\sqrt{\eta_1}\sqrt{\eta_3} \end{cases} \quad (34)$$

where

$$\begin{cases} \eta_1 = \varepsilon_n + \Lambda \\ \eta_2 = 2\Lambda - \chi \\ \eta_3 = \frac{1}{4} + \Lambda - \beta \end{cases} \quad (35)$$

From the knowledge of NU method, we choose the expression $\pi(s)_-$ which the function $\tau(s)$ has a negative derivative. This is given by

$$k_- = -(2\Lambda - \chi) - 2\sqrt{\varepsilon_n + \Lambda}\sqrt{\frac{1}{4} + \Lambda - \beta} \quad (36)$$

with $\tau(s)$ being obtained as

$$\tau(s) = 1 - 2z - 2\left(\sqrt{\varepsilon_n + \Lambda} + \sqrt{\frac{1}{4} + \Lambda - \beta}\right)z + 2\sqrt{\varepsilon_n + \Lambda} \quad (37)$$

Referring to Eq. (13), we define the constant $\lambda$ as

$$\lambda = -(2\Lambda - \chi) - 2\sqrt{\varepsilon_n + \Lambda}\sqrt{\frac{1}{4} + \Lambda - \beta} - \frac{1}{2} - \sqrt{\frac{1}{4} + \Lambda - \beta} - \sqrt{\varepsilon_n + \Lambda} \quad (38)$$

Taking the derivative of $\tau(s)$ from Eq.(37), we have;

$$\tau'(z) = -2 - 2\left(\sqrt{\varepsilon_n + \Lambda} + \sqrt{\frac{1}{4} + \Lambda - \beta}\right) < 0 \quad (39)$$

and $\sigma(z)$ from Eq.(30), we have;

$$\sigma''(z) = -2 \quad (40)$$

Substituting Eq. (39) into Eq. (40), we have

$$\lambda_n = n^2 + n + 2n\sqrt{\varepsilon_n + \Lambda} + 2n\sqrt{\frac{1}{4} + \Lambda - \beta} \quad (41)$$

Comparing Eq. (38) and (41), and carrying out some algebraic manipulation. We have;

$$\varepsilon_n = -\Lambda + \frac{1}{4}\left[\frac{\left(n + \frac{1}{2} + \sqrt{\frac{1}{4} + \Lambda - \beta}\right)^2 - \chi + \beta + \Lambda}{\left(n + \frac{1}{2} + \sqrt{\frac{1}{4} + \Lambda - \beta}\right)}\right]^2 \quad (42)$$

Substituting Eqs. (17) and Eq. (32) into Eq. (31) yields the energy eigenvalue equation of the Hellman potential in the form

$$E_{n\ell} = \frac{\hbar^2\alpha^2\Lambda}{2\mu} - V_1 - \frac{\hbar^2\alpha^2}{8\mu}\left[\frac{\left(n+\frac{1}{2}+\sqrt{\frac{1}{4}+\Lambda-\frac{2\mu V_1}{\hbar^2}}\right)^2 - \frac{4\mu V_1}{\hbar^2\alpha} + \frac{2\mu V_1}{\hbar^2}+\Lambda}{\left(n+\frac{1}{2}+\sqrt{\frac{1}{4}+\Lambda-\frac{2\mu V_1}{\hbar^2}}\right)}\right]^2 \tag{43}$$

The corresponding wave functions can be evaluated by substituting $\pi(s)\_$ and $\sigma(s)$ from Eq. (34) and Eq. (30) respectively into Eq. (7) and solving the first order differential equation. This gives

$$A(z) = z^{\sqrt{\varepsilon_n+\Lambda}}(1-z)^{\frac{1}{2}+\sqrt{\frac{1}{4}+\Lambda-\beta}} \tag{44}$$

The weight function $\rho(s)$ from Eq. (10) can be obtained as

$$\rho(z) = z^{2\sqrt{\varepsilon_n+\Lambda}}(1-z)^{2\sqrt{\frac{1}{4}+\Lambda-\beta}} \tag{45}$$

From the Rodrigues relation of Eq. (9), we obtain

$$y_n(z) \equiv \Omega_{n,l} P_n^{\left(2\sqrt{\varepsilon_n+\Lambda},\, 2\sqrt{\frac{1}{4}+\Lambda-\beta}\right)}(1-2z) \tag{46}$$

where $P_n^{(\theta,\vartheta)}$ is the Jacobi Polynomial.

Substituting $A(s)$ and $y_n(s)$ from Eq. (44) and Eq. (46) respectively into Eq. (5), we obtain the wave function in terms of hyper-geometric polynomial as

$$R_n(z) = \Omega_{n,l}\, z^{\varpi}(1-z)^{\vartheta}\frac{(2\omega+1)_n}{n!}\, {}_2F_1(-n, 2\varpi+\vartheta+n; 2\varpi+1; z) \tag{47}$$

where $\Omega_{n,l}$ is a normalization constant, $\vartheta = \frac{1}{2}+\sqrt{\frac{1}{4}+\Lambda-\beta}$, $\varpi = \sqrt{\varepsilon_n+\Lambda}$ and $(2\varpi+1)_n$ is the Pochhammer's symbol(for the rising factorial)

Using the normalization condition, we obtain the normalization constant as follows [60]:

$$\int_0^\infty R_{n,\ell}(r) \times R_{n,\ell}(r)^* dr = 1 \tag{48}$$

$$-\frac{1}{\alpha}\int_1^0 |R_{n,\ell}(z)|^2 \frac{dz}{z} = 1,\, z = e^{-\alpha r} \tag{49}$$

$$\frac{1}{2\alpha}\int_{-1}^1 |R_{n,\ell}(y)|^2 \frac{2}{1-y}dy = 1,\, y = 1-2z \tag{50}$$

Substituting Eq. (47) into Eq. (50), we have

$$\frac{\Omega_{n\ell}^2}{2\alpha} \int_{-1}^{1} \left(\frac{1-y}{2}\right)^a \left(\frac{1+y}{2}\right)^u \left[P_n^{(a,u-1)}(y)\right]^2 dy = 1, \tag{51}$$

where

$$u = 1 + 2\sqrt{\frac{1}{4} + \Lambda - \beta} \text{ and}$$

$$a = 2\sqrt{\varepsilon_n + \Lambda} \tag{53}$$

Comparing Eq. (51) with the integral of the form [57]

$$\int_{1}^{-1} \left(\frac{1-p}{2}\right)^x \left(\frac{1+p}{2}\right)^y \left[P_n^{(x,y-1)}(p)\right]^2 dp = \frac{2\Gamma(x+n+1)\Gamma(y+n+1)}{n!x\Gamma(x+y+n+1)} \tag{54}$$

We have the normalization constant as

$$\Omega_{n\ell} = \sqrt{\frac{n!a\Gamma(a+u+n+1)}{2\Gamma(a+n+1)\Gamma(u+n+1)}} \tag{55}$$

### 3.2 Solutions of the angular Schrodinger equation for Non-Central Generalized Inverse Quadratic Yukawa Potential

In order to get the solution of equation Eq. (20), we introduce a coordinate transformation of the form,

$z = cos^2\theta$ and Eq. (20) becomes

$$\frac{d^2\Theta(z)}{dz^2} + \frac{(1-3z)}{2z(1-z)} \frac{d\Theta(z)}{dz} + \frac{1}{(2z(1-z))^2} \left(-(\Lambda + B)z^2 + (\Lambda - A - m^2)z - C\right)\Theta(z) = 0 \tag{56}$$

Similarly, Comparing Eq. (56) and Eq. (4), we have the following parameters

$$\begin{cases} \tilde{\tau}(s) = (1 - 3z) \\ \sigma(s) = 2z(1-z) \\ \tilde{\sigma}(s) = -(\Lambda + B)z^2 + (\Lambda - A - m^2)z - C \end{cases} \tag{57}$$

Substituting these polynomials into Eq. (12), we get $\pi(s)$ to be

$$\pi(z) = -\frac{1-z}{2} \pm \sqrt{\left(\frac{1}{4} + (\Lambda + B) - k\right)z^2 + \left(-\frac{1}{2} - (\Lambda - A - m^2) + k\right)z + \frac{1}{4} + C} \tag{58}$$

Further rearranged as;

$$\pi(z) = -\frac{1-z}{2} \pm \sqrt{\left(\frac{1}{4} + \Lambda + B - k\right)z^2 + \left(k - \frac{1}{2} - \Lambda + A + m^2\right)z + \frac{1}{4} + C} \tag{59}$$

To find the constant $k$, the discriminant of the expression under the square root of Eq. (58) should be equal to zero. As such, we have that

$$k_{\pm} = -\frac{(\Lambda - A - m^2 - C)}{2} \pm \frac{1}{2}\sqrt{1 + 4C}\sqrt{C + A + m^2 + B} \tag{60}$$

Substituting Eq. (60) into Eq. (58) yields

$$\pi = -\frac{z}{2} \pm \frac{1}{2}\left(\left(2\sqrt{1 + 4C} + \sqrt{C + A + m^2 + B}\right)z - 2\sqrt{1 + 4C}\right) \tag{61}$$

From the knowledge of NU method, we choose the expression $\pi(s)_-$ which the function $\tau(s)$ has a negative derivative. This is given by

$$k_- = -\frac{(\Lambda - A - m^2 - C)}{2} - \frac{1}{2}\sqrt{1 + 4C}\sqrt{C + A + m^2 + B} \tag{62}$$

with $\tau(s)$ being obtained as

$$\tau(s) = 2 - 4z - 2\left(\sqrt{1 + 4C} + \sqrt{C + A + m^2 + B}\right)z + 2\sqrt{1 + 4C} \tag{63}$$

Referring to Eq. (13), we define the constant $\lambda$ as

$$\lambda = -\frac{(\Lambda - A - m^2 - C)}{2} - \frac{1}{2}\sqrt{1 + 4C}\sqrt{C + A + m^2 + B} - \frac{1}{2} - \frac{1}{2}\left(2\sqrt{1 + 4C} + \sqrt{C + A + m^2 + B}\right) \tag{64}$$

Taking the derivative of $\tau(s)$ from Eq.(63), we have;

$$\tau'(z) = -4 - 2\left(\sqrt{1 + 4C} + \sqrt{C + A + m^2 + B}\right) \tag{65}$$

and $\sigma(z)$ from Eq.(57), we have;

$$\sigma''(z) = -4 \tag{66}$$

Substituting Eq. (65) into Eq. (66), we have

$$\lambda_{\tilde{n}} = 2\tilde{n}^2 + 2\tilde{n} + \tilde{n}\sqrt{1 + 4C} + \tilde{n}\sqrt{C + A + m^2 + B} \tag{67}$$

Comparing Eqs (67) and (64)($\lambda = \lambda_{\tilde{n}}$), and carrying out some algebraic manipulation. We have;

$$\Lambda = \left(2\tilde{n} + 1 + \sqrt{C + A + m^2 + B}\right)^2 + \sqrt{1 + 4C}\left(2\tilde{n} + 1 + \sqrt{C + A + m^2 + B}\right) + C - B \tag{68}$$

or

$$\Lambda = \left(2\tilde{n} + 1 + \sqrt{C + A + m^2 + B}\right)\left(2\tilde{n} + 1 + \sqrt{C + A + m^2 + B} + \sqrt{1 + 4C}\right) + C - B \tag{69}$$

The corresponding wave functions can be evaluated by substituting $\pi(s)_-$ and $\sigma(s)$ from Eq. (57) and Eq. (61) respectively into Eq. (7) and solving the first order differential equation. This gives

$$E(z) = z^{\frac{1}{4}+\frac{1}{2}\sqrt{\frac{1}{4}+C}}(1-z)^{\frac{1}{2}\sqrt{C+A+m^2+B}} \tag{70}$$

The weight function $\rho(s)$ from Eq. (10) can be obtained as

$$\rho(z) = z^{\sqrt{\frac{1}{4}+C}}(1-z)^{\sqrt{C+A+m^2+B}} \tag{71}$$

From the Rodrigues relation of Eq. (9), we obtain

$$y_{\tilde{n}}(z) \equiv \chi_{\tilde{n},m} P_{\tilde{n}}^{\left(\sqrt{\frac{1}{4}+C},\sqrt{C+A+m^2+B}\right)}(1-2z) \tag{72}$$

where $P_n^{(\theta,\vartheta)}$ is the Jacobi Polynomial.

Substituting $E(s)$ $and$ $y_{\tilde{n}}(s)$ from Eq. (70) and Eq. (72) respectively into Eq. (5), we obtain the wave function in terms of hypergeometric Polynomials as;

$$\Theta_{\tilde{n}m}(z) = \chi_{\tilde{n},m} z^{\nu}(1-z)^{\xi} \frac{(2\nu+1)_n}{n!} {}_2F_1(-n, 2\nu+\xi+n; 2\nu+1; z) \tag{73}$$

where $\chi_{\tilde{n},m}$ is a normalization constant, $\nu = \frac{1}{4} + \frac{1}{2}\sqrt{\frac{1}{4}+C}$, $\xi = \frac{1}{2}\sqrt{C+A+m^2+B}$ and $(2\nu+1)_n$ is the Pochhammer's symbol(for the rising factorial)

Now using Eq. (43), we obtain the discrete energy eigenvalues as

$$E_{n\tilde{n},m} = \frac{\hbar^2\alpha^2\kappa}{2\mu} - V_0 - \frac{\hbar^2\alpha^2}{8\mu}\left[\frac{\left(n+\frac{1}{2}+\sqrt{\frac{1}{4}+\kappa-\frac{2\mu V_0}{\hbar^2}}\right)^2 - \frac{4\mu V_0}{\hbar^2\alpha} + \frac{2\mu V_0}{\hbar^2}+\kappa}{\left(n+\frac{1}{2}+\sqrt{\frac{1}{4}+\kappa-\frac{2\mu V_0}{\hbar^2}}\right)}\right]^2 \tag{74}$$

$$\kappa = \left(2\tilde{n}+1+\sqrt{C+A+m^2+B}\right)\left(2\tilde{n}+1+\sqrt{C+A+m^2+B}+\sqrt{1+4C}\right) + C - B \tag{75}$$

where $\tilde{n}$ is the number of nodes of the radial wave functions. The $\Lambda$ is the contribution from the angle-dependent part of the potential and plays the role of centrifugal term.

$$\psi(r,\theta,\phi) = \frac{N_{\tilde{n}m}}{\sqrt{2\pi}} z^{\sqrt{\varepsilon_n+\Lambda}}(1-z)^{\frac{1}{2}+\sqrt{\frac{1}{4}+\Lambda-\beta}} P_n^{\left(2\sqrt{\varepsilon_n+\Lambda},2\sqrt{\frac{1}{4}+\Lambda-\beta}\right)}(1-2z)(\cos^2\theta)^{\frac{1}{4}+\frac{1}{2}\sqrt{\frac{1}{4}+C}}(\sin^2\theta)^{\frac{1}{2}\sqrt{C+A+m^2+B}} P_{\tilde{n}}^{\left(\sqrt{\frac{1}{4}+C},\sqrt{C+A+m^2+B}\right)}(-\cos 2\theta)e^{-im\phi} \tag{76}$$

where $N_{n\,\tilde{n}m}$ is the new normalization constant

## 4 DISCUSSION OF NUMERICAL RESULT

To test the accuracy of our result, we have also computed the energy eigenvalues of the Generalised Inverse Quadratic Yukawa Potential using the energy equation given in Eq. (43).

Our results shown in Table 1 are in good agreement with the result of Ref. [51] who solved the Schrodinger Equation with this potential using the Proper Quantization rule within the framework of the Pekeris Approximation scheme.

Table 1. The bound state energy levels (in units of fm−1) of the GIQYP for various values of $n$, $\ell$ and for $\hbar = \mu = 1$.

| $n$ | $\ell$ | Present $E_{nl}$ $V_1 = 0.5, \alpha = 0.001$ | PQR[51] | Present $E_{nl}$ $V_1 = 1, \alpha = 0.001$ | PQR[51] | Present $E_{nl}$ $V_1 = 0.5, \alpha = 0.01$ | PQR[51] | Present $E_{nl}$ $V_1 = 1, \alpha = 0.01$ | PQR[51] |
|---|---|---|---|---|---|---|---|---|---|
| 0 | 1 | -0.6899105635 | −0.6896035396138025 | -2.995002125 | −2.994004500000000 | -0.6803042120 | −0.6774133455125515 | -2.950212500 | −2.940450000000000 |
| 1 | 1 | -0.5722309407 | −0.5718076188414074 | -1.498001000 | −1.497004500000000 | -0.5658376361 | −0.5619400365376546 | -1.480100000 | −1.470450000000000 |
| 2 | 1 | -0.5375834835 | −0.5371277912043335 | -1.220779125 | −1.219784500000000 | -0.5321979244 | −0.5281909972598103 | -1.207912500 | −1.198450000000000 |
| 3 | 1 | -0.5228774305 | −0.5224100938146438 | -1.123752125 | −1.122760125000000 | -0.5179887921 | −0.5141450483150030 | -1.112712500 | −1.103512500000000 |
| 0 | 2 | -0.5632255334 | −0.5630508543825075 | -1.302587714 | −1.302201872184403 | -0.5547330506 | −0.5533867400061118 | -1.282717604 | −1.279268338661399 |
| 1 | 2 | -0.5340428341 | −0.5337240097533026 | -1.156041314 | −1.155363939225181 | -0.5274476690 | −0.5249118232916035 | -1.141485576 | −1.135367365171019 |
| 2 | 2 | -0.5211298050 | −0.5207493205425896 | -1.094735482 | −1.093938232041968 | -0.5154443911 | −0.5125947736790960 | -1.082471793 | −1.075445608618366 |
| 3 | 2 | -0.5143067798 | −0.5138959792881225 | -1.063404806 | −1.062548531648839 | -0.5091769501 | −0.5063882922242523 | -1.052386409 | −1.045120125270810 |
| 0 | 3 | -0.5327236750 | −0.5326018749170197 | -1.143947977 | −1.143685816860237 | -0.5243377329 | −0.5238335921350012 | -1.125851026 | −1.123949057454857 |
| 1 | 3 | -0.5204455615 | −0.5201906956637058 | -1.088846573 | −1.088310737089324 | -0.5135708753 | −0.5120761646075057 | -1.074282198 | −1.069983262515873 |
| 2 | 3 | -0.5139022862 | −0.5135787736130005 | -1.060095327 | −1.059418943879603 | -0.5079081522 | −0.5061129917794308 | -1.047448733 | −1.042124745183449 |
| 3 | 3 | -0.5100083296 | −0.5096463318898519 | -1.043224516 | −1.042468019323530 | -0.5046167139 | −0.5028802557635820 | -1.031781243 | −1.026091932802665 |
| 0 | 4 | -0.5199924480 | −0.5199025668480347 | -1.085862054 | −1.085664968464422 | -0.5115902204 | −0.5118093030597658 | -1.068266490 | −1.067417988764047 |
| 1 | 4 | -0.5136230846 | −0.5134143288870058 | -1.058372343 | −1.057929861731356 | -0.5065129324 | −0.5059709105589428 | -1.043575818 | −1.040702385388879 |
| 2 | 4 | -0.5098212017 | −0.5095438530724833 | -1.042133812 | −1.041548732718895 | -0.5035609347 | −0.5028023247194076 | -1.029068946 | −1.025236123010328 |
| 3 | 4 | -0.5073721221 | −0.5070530264169021 | -1.031750221 | −1.031076397206152 | -0.5017404815 | −0.5010876689076859 | -1.019873450 | −1.015671686891542 |

## 5. CONCLUSIONS

We have obtained exact bound states solutions of three dimensional three dimensional Schrödinger equation for the Non-Central Generalized Inverse Quadratic Yukawa Potential within the framework of Nikiforov–Uvarov method. We have solved the corresponding eigenvalues and eigenfunctions of the radial and angular parts of the Schrödinger equation. More so, when the ring shaped like term vanishes, i.e. $A = B = C = 0$ then the results are in good agreement with Ref. [51]. Different ring shaped like potentials can be obtained from this new proposed Non central like potential. Finally, our results can find many applications in nuclear physics and quantum chemistry such as cyclic benzene.

## ACKNOWLEDGMENTS

The author dedicates this work to his late father (Mr Okon Edet Udo). He appreciates Dr. A. N. Ikot for communicating some his research materials to him and for continuous encouragement and support.

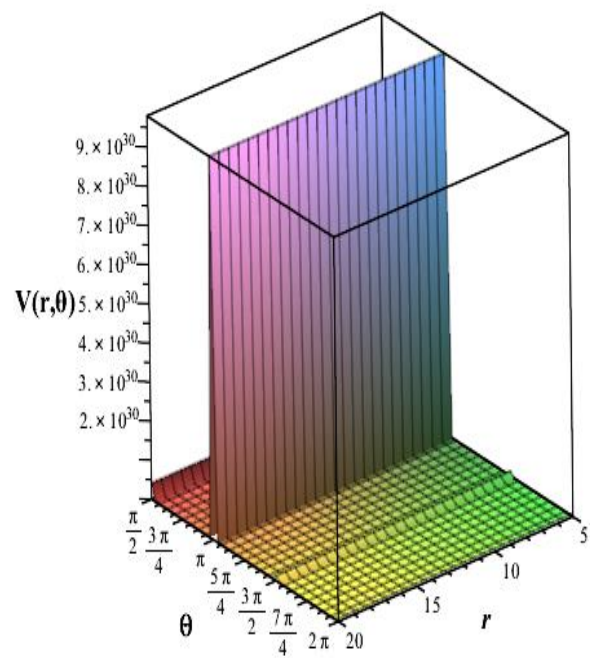

Fig. 1.; Non-central potential for different values of $r$ and $\theta$.